\documentclass[aps, prb, twocolumn, superscriptaddress, noshowpacs, floatfix]{revtex4}
\usepackage{graphicx,epsfig}% Include figure files
\usepackage{amsmath}% bold math
\usepackage{xcolor}% coloured text
\usepackage[breaklinks=true,colorlinks,citecolor=blue,linkcolor=blue,urlcolor=blue]{hyperref}
\def\be{\begin{equation}}
\def\ee{\end{equation}}
\def\nn{\nonumber}
\def\ber{\begin{eqnarray}}
\def\eer{\end{eqnarray}}
\begin{document}
%

%\title{Spin charge coupling in a spin-polarised two dimensional electron gas}
\title{Long-lived spin plasmons in a spin-polarized two-dimensional electron gas}

\author{Amit Agarwal}
\email{amitag@iitk.ac.in}
\affiliation{Department of Physics, Indian Institute of Technology, Kanpur 208016, India}
\author{Marco Polini}
\affiliation{NEST, Scuola Normale Superiore and Istituto Nanoscienze-CNR, I-56126 Pisa, Italy}
\author{Giovanni Vignale}
\affiliation{Department of Physics, University of Missouri, Columbia, Missouri 65211, USA}
\author{Michael E. Flatt\'{e}}
\email{michael\_flatte@mailaps.org}
\affiliation{Department of Physics and Astronomy and Optical Science and Technology Center, University of Iowa, Iowa City, Iowa 52242, USA}
\begin{abstract}
Collective charge-density modes (plasmons) of the clean two-dimensional unpolarized electron gas are stable, for momentum conservation prevents them from decaying into single-particle excitations. Collective spin-density modes (spin plasmons) possess no similar protection and rapidly decay by production of electron-hole pairs.  Nevertheless, if the electron gas  has a sufficiently high degree of spin polarization ($P>1/7$, where $P$ is the ratio of the equilibrium spin density and the total electron density, for a parabolic single-particle spectrum) we find that a long-lived spin-plasmon---a collective mode in which the densities of up and down spins oscillate with opposite phases---can exist within a ``pseudo gap'' of the single-particle excitation spectrum.  The ensuing collectivization of the spin excitation spectrum is quite remarkable and should be directly visible in Raman scattering experiments.  The predicted mode could dramatically improve the  efficiency of coupling between  spin-wave-generating devices, such as spin-torque oscillators. 
\end{abstract}
% 
%\pacs{}
%
\maketitle
\section{Introduction}
\begin{figure}[t]
\begin{center}
\includegraphics[width=.98 \linewidth]{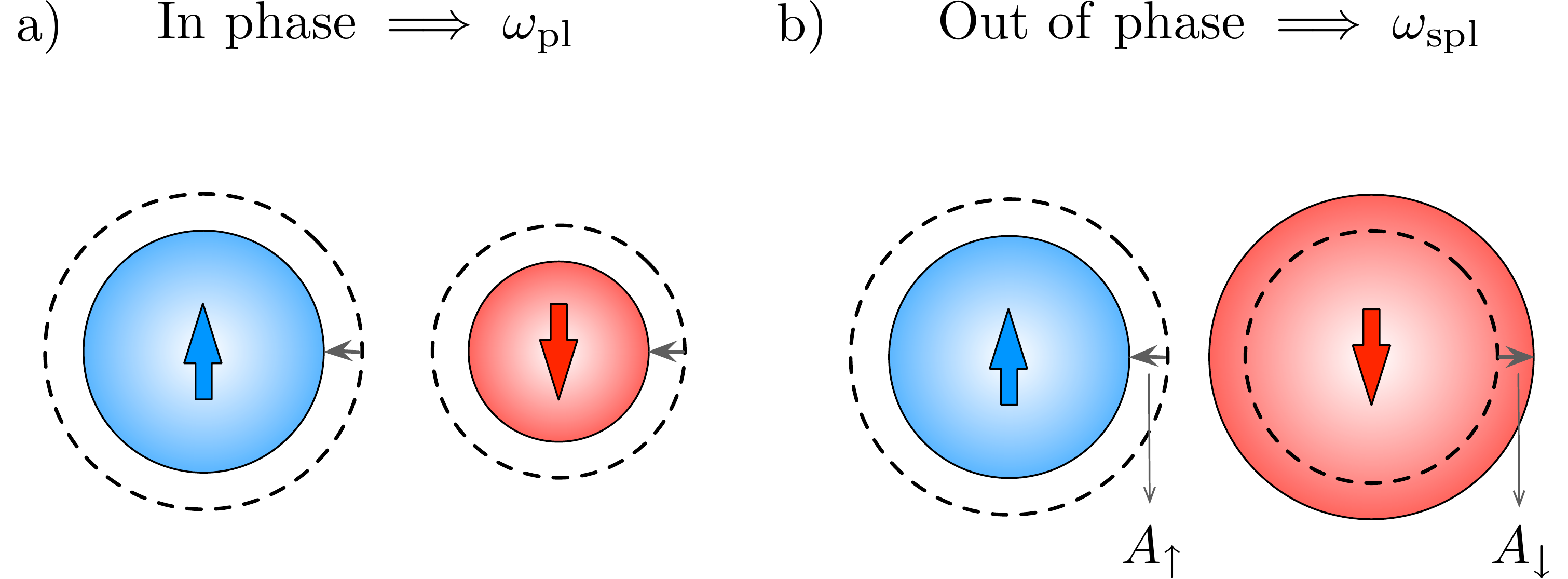}
\caption {Cartoon depicting that a) the in phase oscillations of the two spin fluids give rise to a charge plasmon, and b) the out of phase oscillations of the two spin fluids gives rise to the spin plasmon. The ratio of the oscillation amplitudes of the two spin fluids, for the case of the spin plasmon mode in panel b), is given by Eq.~\eqref{eq:ratio}~. The dashed and the colored circles  (blue for $\uparrow$-spin and red for $\downarrow$-spin) represent the equilibrium and displaced densities of the two spin fluids, respectively in both the panels. 
\label{fig:fig0}}
\end{center}
\end{figure}

The strong electromagnetic fields localized near metallic nanoparticles, the deceleration of a charged particle in a metal, and the dramatic reduction in effective wavelength of electromagnetic waves near a metal surface are all associated with charge plasmons, the stable charge-density collective modes of an electronic system \cite{Pines, Giuliani_and_Vignale, Maier, Ebbesen}. Stable collective spin modes of ferromagnetic systems (magnons) have also been extensively explored due to their role in reducing the Curie temperature of the ferromagnet, their effect on ferromagnetic resonance linewidths and quality factors in microwave devices\cite{spinwavebook}, and their potential role in efficient low-energy information transfer\cite{Kajiwara2010nature,Lenk2011_PhysRep,Kruglyak2010_JPD,Khitun2010_JPD}. The stability of both of these types of modes occurs because the momentum and energy of a mode does not overlap with the single-particle excitations of the system, so the linewidth of the mode is substantially less than the mode energy. By comparison the collective spin modes in nonmagnetic systems (spin plasmons) have drawn much less attention, due to the expectation that those modes will rapidly decay into single-particle states. However, the situation readily changes when the electron gas is spin-polarized---a state of affairs that can be achieved by various means,  including spin injection \cite{Ohno1}, optical excitation\cite{optical-orientation} and current-induced spin polarization\cite{current}.
%, or by applying {\textcolor{blue} {an in-plane magnetic field (thus avoiding problems with the orbital coupling)}. 
Then, in addition to the well-known magnons\cite{Kittelbook}, which are simply oscillations in the {\it direction} of the spin polarization,  a long-lived spin plasmon mode emerges.  In this mode, the densities of up and down spins oscillate with opposite phases  [see Fig.~\ref{fig:fig0}], causing the spin polarization to change in magnitude but not in direction.
%%%%%%%%%%%% 
%%%%%%%%%%%%%%%%%%%%%%%%%%%%%%%%%%%%%%%%%%%%%%%%%%%
%%%%%%%%%%%% 
\begin{figure}[t]
\begin{center}
\includegraphics[width=1.0 \linewidth]{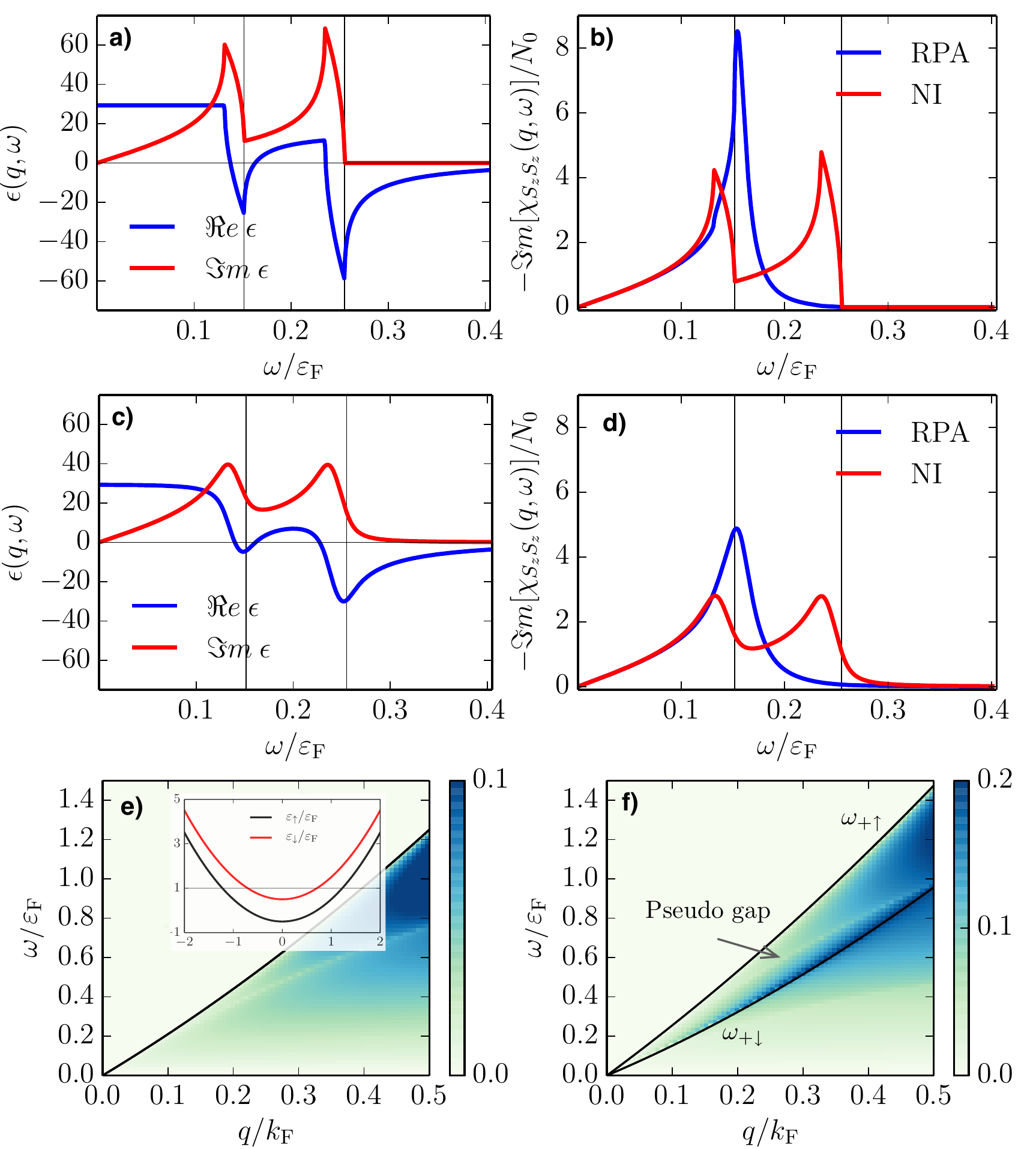}
\caption {Panel a) shows the real and imaginary parts of the RPA dielectric function $\epsilon(q,\omega)$ versus $\omega$ for a clean system.  Panel b) displays $\Im m [\chi_{S_zS_z}(q,\omega)]$ versus $\omega$ for a clean system, with and without Coulomb interactions. Panel c) and panel d) show the real and imaginary parts of the dielectric function versus $\omega$ and  $\Im m[\chi_{S_zS_z}(q,\omega)]$ versus $\omega$, respectively, for a disordered system with $(\tau \varepsilon_{\rm F})^{-1} = 0.01$.  Other parameters are chosen to be $P = 0.5$, $q/k_{\rm F}= 0.1$, and $r_s =2$. The left and right vertical lines in the plots mark the upper limit of the electron-hole continuum, of the minority ($\omega_{+ \downarrow}$ for $P>0$) and majority spin species ($\omega_{+ \uparrow}$ for $P>0$), respectively. Note that the spin plasmon mode arises from the second zero of the dielectric function (from the origin), in panels a) and c), and it always lies between 
$\omega_{+\downarrow}$ and $\omega_{+\uparrow}$. In panels b) and d) ``NI'' stands for ``non-interacting''. Panels e) and f) display the colour plot of the loss function [$-1/\epsilon(q,\omega)$] in the $q-\omega$ plane for $P=0$ and $P=0.5$ respectively. The inset in panel e) shows the spin split energy bands. 
\label{fig:fig1} }
\end{center}
\end{figure}
%%%%%%%%%%%%%%%%%%%%%%%%%%%%%%%%%%%%%%%%%%%%%%%%%%%

The  physics of spin plasmons in the spin-polarized two-dimensional electron gas (SP2DEG) is easily understood from the dielectric function, shown in Figs.~\ref{fig:fig1}a) and~b).  The red line in Fig.~\ref{fig:fig1}a), corresponding to $\Im m [\epsilon(q,\omega)]$,  is the non-interacting single-particle excitation spectrum scaled by the strength of the Coulomb interaction; the two peaks correspond to excitations from the Fermi sea of majority spins (high-frequency peak) and minority spins (low-frequency peak).  A region of reduced spectral density, which we refer to as a ``pseudo gap" is clearly visible between the two, just above the  the minority-spin peak.  The blue line is $\Re e[ \epsilon(q,\omega)]$, calculated in the random phase approximation (RPA)~\cite{Pines, Giuliani_and_Vignale}, which is essentially exact in the high-density limit.   The zeroes of $\Re \epsilon(q,\omega)$, if they occur in a region of sufficiently low density of single-particle excitations,  indicate the possibility of spontaneous, self-sustained, collective oscillations of the electronic densities.  Such a zero is clearly seen to be present in the pseudo-gap region of Fig.~\ref{fig:fig1}a). 

Fig.~\ref{fig:fig1}b) shows that  the two-peaked non-interacting spectrum (red line, same as Fig.~\ref{fig:fig1}a) is replaced by a single peak centered at the frequency of the spin plasmon in the interacting excitation spectrum (blue line), which is calculated in the RPA.
This ``collectivization" of the spectrum should be observable in Raman scattering. These features are robust to disorder, as shown in Fig.~\ref{fig:fig1}cd) for a disordered system with the carrier scattering time $\tau$ such that $(\tau \varepsilon_{\rm F})^{-1} = 0.01$.
In the long-wavelength limit the mode is charge-neutral --- the displacement of up spins compensates an opposite displacement of down spins --- and the equilibrium-restoring Coulomb force vanishes, resulting in a vanishing frequency for $q \to 0$.  At finite values of $q$ the majority up-spins no longer exactly compensate the minority down-spins; a small charge component appears, resulting in a finite restoring force and a frequency that is proportional to $q$.

In this Article we present a complete semi-analytic theory of spin plasmon dispersion and damping in a SP2DEG, in the presence of weak disorder.  While the possibility of spin plasmons has been recognized in earlier work~\cite{Marinescu,Ryan,Perez,Magarill,Agarwal}, this is to the best of our knowledge the first time that the crucial issue of the robustness of the mode with respect to electron-hole pair generation and disorder is addressed.  Our conclusion is that the spin plasmon is indeed robust against all these potentially destabilizing effects,  as well as Fermi liquid corrections (beyond RPA), which become relevant at low density.  We further suggest that spin plasmons could be used as an effective means of coupling spintronic devices such as spin-torque oscillators, which produce localized spin-wave excitations: by tuning the spin polarization of the SP2DEG the coupling could be altered from efficient (when the plasmon mode is stable, and the mode lies outside the single-particle continuum of minority-spin excitations) to inefficient (when the plasmon mode quickly decays and the mode is within this continuum).%

Our Article is organized as follows.  In Section~\ref{sect:RPAtheory} we present the RPA theory of the spin plasmon in a SP2DEG.  Section~\ref{sect:analyticalresults}, in particular, presents our analytical results for the clean limit, while in Section~\ref{sect:mermin} we address the important issue of the effect of disorder through Mermin's relaxation time approximation. In Section~\ref{sect:Fermiliquidcorrections} the effect of Fermi liquid corrections beyond the RPA is briefly discussed. Section~\ref{sect:summary} presents a brief summary of our main conclusions.

\section{RPA theory of spin plasmons in SP2DEGs}
\label{sect:RPAtheory}

We consider a zero-temperature SP2DEG with the usual parabolic-band dispersion relation and Fermi energy $\varepsilon_{\rm F}$. A spin polarization could be induced through various means, including spin injection, optical excitation, and current-induced spin polarization. However, for the sake of simplicity, we here assume that it is induced by an in-plane Zeeman field $B$, which has negligible orbital effects when the width of the quantum well hosting the 2D electronic system is sufficiently small. The spin-resolved bands have energy $\varepsilon_{\uparrow(\downarrow)}(k) = \hbar^2 k^2 /(2 m_{\rm b}) \pm g_{\rm b} \mu_{\rm B} B  $ and the spin polarization is $P\equiv (n_\uparrow - n_\downarrow)/(n_\uparrow + n_\downarrow) = - g_{\rm b} \mu_{\rm B} B/\varepsilon_{\rm F}$ for small magnetic fields. Here $m_{\rm b}$ and $g_{\rm b}$ are the band mass and the Land\'{e} g-factor, respectively, while $\mu_{\rm B}$ is the Bohr magneton. The spin-resolved Fermi wave vectors $k_{\rm F \uparrow (\downarrow)}$ can be conveniently written in terms of the spin polarization $P$ as $k_{\rm F \uparrow (\downarrow)} = k_{\rm F} \sqrt{1\pm P} $, where we have defined $\hbar k_{\rm F} \equiv \sqrt{2 m_{\rm b} \varepsilon_{\rm F}} $. Identical expressions hold for the spin-resolved Fermi velocities $v_{\rm F \uparrow (\downarrow)}$ in terms of  $v_{\rm F} \equiv \hbar k_{\rm F}/m_{\rm b}$.  
The total electron density $n$, for convenience, is expressed in terms of the dimensionless Wigner-Seitz parameter~\cite{Giuliani_and_Vignale} $r_s$, which is given by $r_s = (\pi n a^2_{\rm B})^{-1/2}$, where $a_{\rm B} = \epsilon \hbar^2/(m_{\rm b} e^2)$ is the material's Bohr radius, $\epsilon$ being the static dielectric constant of the material hosting the SP2DEG.

Following Ref.~\onlinecite{Giuliani_and_Vignale}, the spin resolved response functions $\chi_{\sigma \sigma'} (q,\omega)$ of the interacting SP2DEG are given in the RPA by the following equation 
\be \label{eq: RPA1}
\begin{pmatrix}
{\chi }_{\uparrow \uparrow} & {\chi }_{\uparrow \downarrow} \\
{\chi }_{\downarrow \uparrow} & {\chi }_{\downarrow \downarrow}
\end{pmatrix}^{-1} = 
\begin{pmatrix}
 \chi^{(0)}_{\uparrow } & 0 \\
0 &  \chi^{(0)}_{\downarrow}
\end{pmatrix}^{-1}  - v_q 
\begin{pmatrix}
1& 1 \\
1 & 1
\end{pmatrix}~,
\ee
where $v_q = 2\pi e^2/(\epsilon q)$  is the 2D Fourier transform of the Coulomb interaction and  $\chi^{(0)}_{\uparrow (\downarrow) } $ is the spin-resolved non-interacting density-density (Lindhard) response function. 
In writing the explicit functional dependence of $v_q$ on $q$ we have omitted a form factor $0 < F (q) < 1$, which arises~\cite{Ando} due to the finite 
width of the quantum well hosting the SP2DEG. The role of $F(q)$ is to weaken the bare Coulomb interaction at wave vectors $q \sim 1/L$ where $L$ is the width of the quantum well. In the long-wavelength limit $F(q \to 0) = 1$. 
Eq.~\eqref{eq: RPA1} is physically transparent: coupling between the spin-up and spin-down electronic subsystems originates from the Coulomb interaction $v_q$, which acts identically within one subsystem and between the two different spin sub-systems.

Solving Eq.~(\ref{eq: RPA1}) for $\chi_{\sigma \sigma'} (q,\omega)$ and changing the response-function basis to total density $n \equiv n_{\uparrow} + n_{\downarrow}$ and total spin $S_z \equiv n_{\uparrow} - n_{\downarrow}$, we obtain the following three response functions:
\be\label{eq:CC}
{\chi }_{nn}(q,\omega) = \frac{S(q,\omega)}{\epsilon(q, \omega)}~,
\ee
\be\label{eq:SS}
{\chi }_{S_z S_z}(q,\omega) = \frac{S(q,\omega) - 4 v_q P(q,\omega)}{\epsilon(q, \omega)}~,
\ee
and
\be\label{eq:SC}
{\chi }_{n S_z} = \frac{D(q,\omega)}{\epsilon(q, \omega)}~,
\ee
where $S(q,\omega) = \chi^{(0)}_{\uparrow}(q,\omega) + \chi^{(0)}_{\downarrow}(q,\omega)$, $D(q,\omega) = \chi^{(0)}_{\uparrow}(q,\omega) - \chi^{(0)}_{\downarrow}(q,\omega)$, $P(q,\omega) = \chi^{(0)}_{\uparrow}(q,\omega) \chi^{(0)}_{\downarrow}(q,\omega)$, and, finally,
\be
\epsilon(q, \omega) \equiv  1- v_q [\chi^{(0)}_{\uparrow}(q,\omega) + \chi^{(0)}_{\downarrow}(q,\omega)]~. 
\ee
In obtaining these results we have used the following identities:
$ {\chi }_{\downarrow \uparrow}  = {\chi }_{\uparrow \downarrow}$,
$
{\chi }_{nn} = {\chi }_{\uparrow \uparrow} + {\chi }_{\downarrow \downarrow} + 2 {\chi }_{\uparrow \downarrow}
$,
$
{\chi }_{S_z S_z} = {\chi }_{\uparrow \uparrow} + {\chi }_{\downarrow \downarrow}  - 2 {\chi }_{\uparrow \downarrow}
$,
and, finally, 
$
{\chi }_{n S_z} = {\chi }_{\uparrow \uparrow} - {\chi }_{\downarrow \downarrow} = {\chi }_{S_z n}
$.
We have also neglected retardation effects due to the finite velocity of propagation of the electromagnetic field.  These effects are negligible at wave vectors $q>\omega_{\rm p}/c$, which implies $q/k_{\rm F}>k_{\rm F}e^2/mc^2\simeq 10^{-8}$.  This condition is well satisfied throughout the range of our calculations.

Collective (plasmon) modes emerge as poles of the response functions (\ref{eq:CC})-(\ref{eq:SC}). Since $S(q,\omega)$, $D(q,\omega)$, and $P(q,\omega)$ are smooth functions of momentum and energy, collective modes coincide with the zeroes of the complex longitudinal ``dielectric function" $\epsilon(q,\omega)$, {\it i.e.}
\be \label{eq:condition}
\epsilon(q, \omega) =  0~.
\ee
The real frequency $\omega_{\rm coll}$ and inverse lifetime $\gamma_{\rm coll}$ (or damping rate) of the collective excitations,  for a given $q$, are obtained from the complex roots of Eq.~\eqref{eq:condition}:  $\omega = \omega_{\rm coll}(q) - i \gamma_{\rm coll}(q)$.  
Note that $\gamma_{\rm coll}>0$ is essential for the stability of the collective mode. The damping of the collective mode typically occurs if the frequency of the collective mode lies in the electron-hole continuum. In this case 
the mode decays by creating single electron-hole pairs (Landau damping). 

An estimate of the damping rate $\gamma_{\rm coll}$, if it is small, can be obtained~\cite{Giuliani_and_Vignale} by doing a Laurent-Taylor  expansion of the dielectric function around $\omega_{\rm coll}$, to obtain (for a frequency independent $v_q$),
\be \label{eq:gamma}
\gamma_{\rm coll}(q) =  \left. \frac{\Im m [\epsilon(q,\omega)]}{\partial \Re e[\epsilon (q,\omega)]/\partial \omega} \right|_{\omega = \omega_{\rm coll}}  ~.    
\ee
In a realistic experimental scenario, the observability of a damped collective mode depends on the sharpness of its resonance peak, which is directly measured by the quality factor (QF) defined as  $Q \equiv \omega_{\rm coll}(q)/\gamma_{\rm coll}(q)$. 
We note that we can safely neglect the spin-relaxation mechanism here, since  spin-relaxation occurs over much larger timescales (of the order of nanoseconds in semiconductors), as compared  to the timescale of plasmon dynamics (TeraHertz).

In the rest of this Article we will focus on the collective excitations in a SP2DEG and show that in addition to a regular plasmon mode (charge density excitations), there is a 
spin-plasmon mode, which emerges from the coupling of spin and charge degrees of freedom in the presence of Coulomb interactions. We emphasize that, within the RPA, the coupling of 
spin and charge degrees of freedom arises only from electrostatic effects, as exchange interactions are not included in the RPA.

\subsection {Spin plasmon mode in the clean limit}
\label{sect:analyticalresults}

After straightforward algebraic manipulations we find that Eq.~(\ref{eq:condition}) allows two solutions in the long-wavelength limit: an undamped charge plasmon with dispersion
\be\label{eq:chargeplasmon}
\omega_{\rm pl}^2(q\to 0)  = \frac{2 \pi n e^2}{m_{\rm b} \epsilon_0} q + {\cal O}(q^2)~,
\ee
and a partially damped, acoustic spin-plasmon mode whose long-wavelength dispersion is,
\be \label{eq:plasmon_ac}
\omega_{\rm spl}(q \to 0 ) =c_{\rm s} q  + {\cal O}(q^2)~,
\ee 
where
\be \label{eq:velocity}
c_{\rm s} = \frac{2}{\sqrt{3}}v_{\rm F}\sqrt{1-P}~,
\ee
which lies between $ v_{\rm F \downarrow}$ and $v_{\rm F \uparrow}$. 
The charge-plasmon mode in Eq.~(\ref{eq:chargeplasmon}) has exactly the same long-wavelength form as that of the plasmon mode in an {\it unpolarized} 2DEG~\cite{Giuliani_and_Vignale,czachor_prb_1982}. 

The spin plasmon  in the SP2DEG arises due to the second zero of the real part of the dielectric function --- see Fig. \ref{fig:fig1}a) --- which lies in a ``pseudo gap'' between the electron-hole continuum of the minority ($\downarrow$-spin for $P>0$) and majority ($\uparrow$-spin for $P>0$) spin species. Spin polarization separates the electron-hole spectrum of the majority and minority spins,
creating a region of low-density electron-hole pairs, just above the minority-spin electron-hole continuum --- see Fig. \ref{fig:fig1}b).  It is  this region of reduced electron-hole density that supports the spin plasmon  mode. As a consequence, the damping of the spin-plasmon mode is never exactly zero, but it can be quite small leading to an acoustic  spin-plasmon mode with a relatively high QF. 

%%%%%%%%%%%%%%%%%%%%%%%%%%%%%%% Figure 2 %%%%%%%%%
\begin{figure}[t]
\begin{center}
\includegraphics[width=1.0 \linewidth]{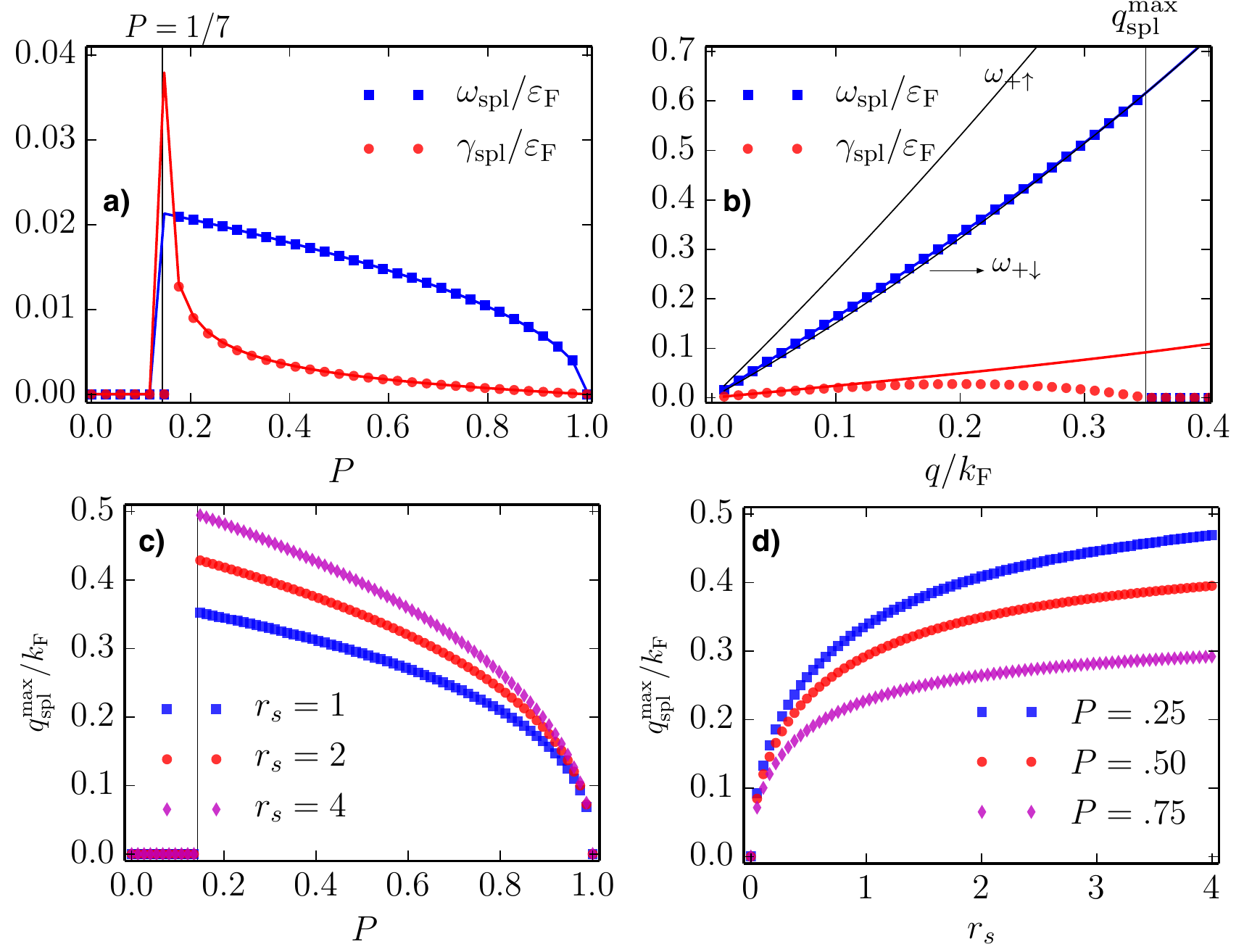}
\caption{ Panels a) and b) display the frequency $\omega_{\rm spl}$ and the damping rate $\gamma_{\rm spl}$ of the spin-plasmon mode as a function of spin polarization and wave vector, respectively, 
for $r_s =2$ and in the absence of disorder. The  points, filled blue squares for $\omega_{\rm spl}$ and filled red circles for $\gamma_{\rm spl}$, are obtained from the numerical calculations, and the respective solid lines (blue and red) are from the analytical expressions [Eq.~\eqref{eq:wac_exact} for $\omega_{\rm coll}$  for all $q$ and Eq.~\eqref{eq:gamma} for $\gamma_{\rm coll}$ in the $q \to 0$ limit].  In panel a) $q/k_{\rm F}=0.01$, 
and the thin vertical line marks the $P=1/7$ line, for visual aid. Note that in Panel b), the spin-plasmon mode ceases to exist beyond a certain $q^{\rm max}_{\rm spl}$ (depicted as a thin vertical line), which is given by the condition: $\omega_{\rm spl}  > \omega_{+\downarrow} $. Here we have chosen $P = 0.5$. Panels c) and d) shows the dependence of $q^{\rm max}_{\rm spl}$ on $P$ and $r_s$, respectively.
 \label{fig:fig2}}
\end{center}
\end{figure}
%%%%%%%%%%%%%%%%%%%%%%%%%%%%%%%%%%%%%%%%%%%%%%%%

The spin-plasmon dispersion can be calculated exactly within RPA by noting that 
in the pseudo gap, for $P>0$ we have $\omega_{+ \downarrow} < \omega < \omega_{+ \uparrow}$, where $\omega_{+\sigma} \equiv q^2/2m + v_{\rm F \sigma}q$ is the upper boundary of the electron-hole continuum for the respective spin species.  In this regime of interest, the real part of the density-density response function\cite{Giuliani_and_Vignale, Ando} of the majority spins ($\uparrow$-spin for $P>0$) is a constant independent of $q$ and $\omega$:  $\Re e[\chi_{0\uparrow}(q,\omega)] = -N_0$, where $N_0 \equiv m/{2 \pi \hbar^2}$ is the 2D density-of-states per spin and per unit volume, at the Fermi surface. 
 The density-density response function of minority spins ($\downarrow$-spin for $P>0$) is purely real:
\be \label{eq: chi0down}
- \frac{\chi^{(0)}_\downarrow}{N_0} = 1+ \frac{k_{{\rm F} \downarrow}}{q} \left(\sqrt{\nu_{- \downarrow}^2 -1} - \sqrt{\nu_{+ \downarrow}^2 -1}~ \right)~,
\ee
where
\be  
\nu_{\pm \sigma} \equiv \frac{\omega + i \eta }{v_{\rm F \sigma}} \pm \frac{q}{2 k_{\rm F \sigma}} ~;~~~(\eta \to 0^{+})~.
\ee
Setting $\Re e [\epsilon(q,\omega)]=0$ yields the algebraic equation
\be \label{eq:dis_condition}
1+V_q\frac{k_{{\rm F}\downarrow}}{q} \left(\sqrt{\nu_{- \downarrow}^2 -1} - \sqrt{\nu_{+ \downarrow}^2 -1} \right)=0~,
\ee
where we have introduced the dimensionless quantity
\be
V_q \equiv \frac {v_qN_0}{1+2v_qN_0}~. 
\ee
Note that $V_q \to 1/2$ for $q \to 0$.
The solution of Eq.~\eqref{eq:dis_condition} is
\be \label{eq:wac_exact}
\frac{\omega_{\rm spl}}{qv_{{\rm F}\downarrow}}=\sqrt{\frac{1}{1-V_q^2}+\frac{q^2}{4k_{{\rm F}\downarrow}^2V_q^2}}~,
\ee
which, in the limit of $q \to 0$, yields Eqs.~\eqref{eq:plasmon_ac}-\eqref{eq:velocity}. 
We note that for our treatment to be valid in the $q \to 0$ limit the condition $\omega < q v_{{\rm F} \uparrow}$ must also be satisfied, and our result in Eq.~\eqref{eq:wac_exact} is compatible with this condition only if
\be
\frac{v_{{\rm F}\uparrow}}{v_{{\rm F}\downarrow}}>\frac{2}{\sqrt{3}}~,~~{\rm or},~~~ P>1/7~.
\ee
The lower bound on the polarization $P$ for the existence of a spin plasmon mode in a SP2DEG, {\it i.e.}~$P > 1/7$, is also evident from Fig. \ref{fig:fig2}a).

In addition to the above condition, for the spin-plasmon mode to lie in the pseudo-gap region an additional condition must be met for finite values of the wave vector $q$. It is $\omega_{\rm spl} > \omega_{+ \downarrow}$: 
this gives us a maximum wave vector $q^{\rm max}_{\rm spl}$ beyond which the spin-plasmon mode ceases to exist---see Fig.~\ref{fig:fig2}b). 
Straightforward algebraic manipulations yield
\be
\frac{q^{\rm max}_{\rm spl}}{k_{\rm F}}= -\frac{4 r_s} {3 \sqrt{2}} + \frac{7 r_s^2}{6 \zeta^{1/3}} + \frac{1}{3} \zeta^{1/3}~,
\ee
where
\ber \label{eq:zeta}
\zeta &= &  -\frac{5}{\sqrt{2}}~ r_s^3 + \frac{27 \sqrt{1-P}}{2}~ r_s^2+3 \sqrt{3}~ r_s^2 \nn \\
& \times& \sqrt{\frac{27 (1-P)}{4} -\frac{5  \sqrt{(1-P)}}{\sqrt{2}}~ r_s -\frac{9}{8} ~r_s^2}~.
\eer
The dependence of $q^{\rm max}_{\rm spl}$ on $P$ and $r_s$ is shown in Figs.~\ref{fig:fig2}c) and~d), respectively.

The damping rate for the spin-plasmon mode, $\gamma_{\rm spl}(q)$, can be calculated from Eq.~\eqref{eq:gamma}. 
In the pseudo-gap region,  the imaginary part of the dielectric function is controlled by the majority-spin electrons:
\ber 
\Im m [\epsilon(q,\omega)] &=& -v_q\Im m[\chi^{(0)}_{\uparrow}(q,\omega)] \\
&=&v_qN_0\frac{k_{{\rm F}\uparrow}}{q} \left(\sqrt{1-\nu_{- \uparrow}^2 } - \sqrt{1-\nu_{+ \uparrow}^2} \right)~. \nn
\eer
Substituting $\omega = \omega_{\rm spl} = 2 v_{{\rm F} \downarrow} q/\sqrt{3}$ in the above equation, we calculate the leading order term in the long wavelength $q \to  0$ limit, to be
\be \label{eq:Imepsilon}
\Im m [\epsilon(q \to 0,\omega\to \omega_{\rm spl})] = v_qN_0 \sqrt{\frac{4(1-P)}{7P-1}} + {\cal O}(q^2)~.
\ee
In the pseudo-gap regime the derivative of the real part of $\epsilon$ with respect to frequency is given by
\be 
\partial_\omega \Re e [\epsilon(q,\omega)] = - v_q \partial_\omega \chi^{(0)}_{\downarrow} (q,\omega)~.
\ee
Note that $\partial_\omega \Re e \epsilon(q,\omega)$ is the inverse of the oscillator strength of the collective mode.
Using Eq.~\eqref{eq: chi0down} in the above equation, we find
\be \label{eq:strength}
\left. \partial_\omega \Re e [\epsilon(q \to 0,\omega)]\right|_{\omega_{\rm spl}} = 3\sqrt{3} ~\frac{v_qN_0}{qv_{{\rm F}\downarrow}} +  {\cal O}(q^0)~ .
\ee
Using Eqs.~\eqref{eq:Imepsilon} and~\eqref{eq:strength}~in Eq.~\eqref{eq:gamma} gives the damping rate or the imaginary part of the spin-plasmon frequency
 \be \label{eq:gamma_spin_plasmon}
 \gamma_{\rm spl}(q\to 0)= \omega_{\rm spl} \frac{1}{3}\sqrt{\frac{1-P}{7P-1}}  + {\cal O}(q^2)~,
 \ee
 and thus the quality factor of the spin-plasmon mode  
 \be
Q_{\rm spl}(q\to 0) \equiv  \frac{\omega_{\rm spl}(q\to 0)}{\gamma_{\rm spl}(q\to 0)}=3\sqrt{\frac{7P-1}{1-P}} + {\cal O}(q)~.
 \ee
We emphasize here that as $P \to 1$, the frequency of the spin-plasmon mode itself vanishes, {\it i.e.}~the spin plasmon mode ceases to exist. Thus the quality factor for $P=1$ here is misleading and the infinity is arising since both the frequency and damping factor are zero. 

Finally we note that the spin-plasmon eigenmode is characterized by oscillations of up and down spin densities with amplitudes  $A_\uparrow (q)$ and $A_\downarrow(q)$ respectively.  These amplitudes satisfy the eigenvalue equation [see Eq.~\eqref{eq: RPA1}],
 \be
 \left(\begin{array}{cc} \Re e[1/\chi^{(0)}_{\uparrow}] - v_q &-v_q\\
 -v_q& \Re e [1/\chi^{(0)}_{\downarrow}]-v_q \end{array}\right)\left(\begin{array}{c} A_\uparrow(q)\\A_\downarrow(q)\end{array}\right)=0~.
 \ee
 The first row of this equation yields
 \be
 \left\{\Re e [1/\chi^{(0)}_{\uparrow}(q,\omega_{\rm spl})] - v_q\right\}A_\uparrow(q) - v_qA_\downarrow(q)=0~.
 \ee
 Making use of the fact that  $\Re e [1/\chi^{(0)}_{\uparrow}(q,\omega_{\rm spl})] = - 1/N_0$ in the regime of existence of the spin-plasmon mode, 
 we find
\be \label{eq:ratio}
 \frac{A_\uparrow}{A_\downarrow}=-\frac{v_qN_0}{1+v_qN_0}~.
\ee
In the limit $q \to 0$ this tends to $-1$, meaning that the densities of up and down spins oscillate with equal amplitudes and opposite phases.
The spin-plasmon mode, which arises due to the interplay of spin-polarized bands and Coulomb interactions, physically corresponds to an excitation in which spin-up and spin-down fluids oscillate 
with equal amplitude but opposite phase.

%%%%%%%%%%%%%%%%%%%%%%%%%%%%%%% Figure 3 %%%%%%%%%
\begin{figure}[t]
\begin{center}
\includegraphics[width=1.0 \linewidth]{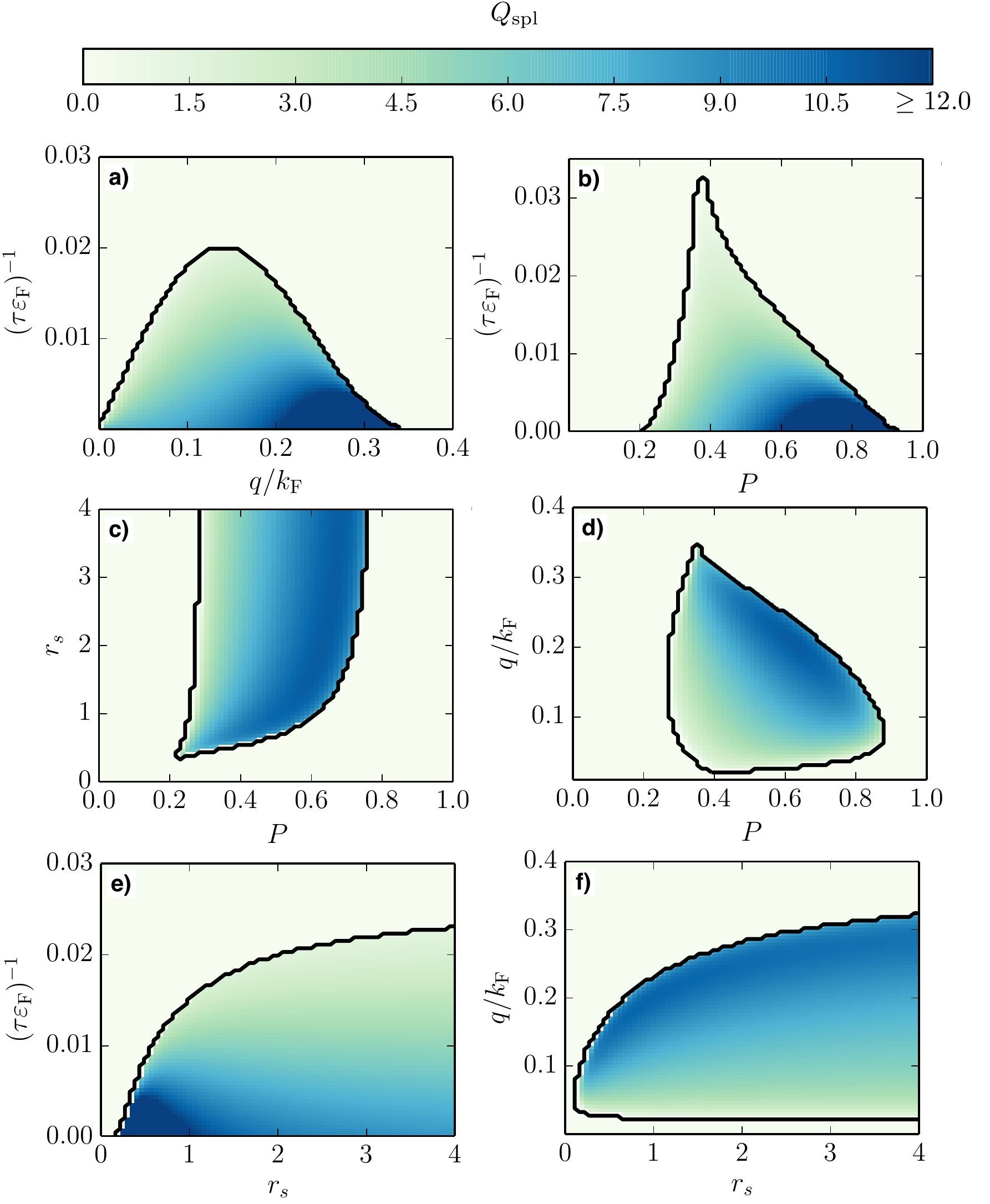}
\caption{Quality factor of the spin-plasmon mode in the presence of disorder. Panel a) shows the quality factor of the spin-plasmon mode, in the $\tau^{-1}-q$ plane for $r_s =2$ and $P = 0.5$. Panel b) illustrates the quality factor of the spin-plasmon mode in the $\tau^{-1}-P$ plane for $r_s =2$ and $q/k_{\rm F} = 0.15$.  Panel c) displays the quality factor of the spin-plasmon mode in the $r_s-P$ plane for $q/k_{\rm F}  =0.15$ and $(\tau \varepsilon_{\rm F})^{-1}  = 0.005$. Panel d) shows the quality factor of the spin-plasmon mode in the $q-P$ plane for $r_s =2$ and $(\tau \varepsilon_{\rm F})^{-1}  = 0.005$. In panel e) we present results for the quality factor of the spin-plasmon mode in the $\tau^{-1}-r_s$ plane for $q/k_{\rm F}  =0.15$ and $P = 0.5$. Finally, panel f) shows the quality factor in the $q-r_s$ plane for  $(\tau \varepsilon_{\rm F})^{-1}  = 0.005$ and $P = 0.5$. In all the panels, the solid black line marks the $Q_{\rm spl} = 0$ contour line, {\it i.e.}~the boundary of existence of the spin-plasmon mode.  Note that we have chosen the upper limit of the color scale to be $12$ to ensure uniformity of all the panels, and to improve the contrast of the figures. The Quality factor is not limited by it and is much higher in some regions.\label{fig:fig3}}
\end{center}
\end{figure}
%%%%%%%%%%%%%%%%%%%%%%%%%%%%%%%%%%%%%%%%%%%%%%%%

%
\subsection{The impact of disorder on spin plasmons in SP2DEGs}
\label{sect:mermin}

We now turn to a discussion of the impact of disorder, provided by impurities, on the collective charge and spin-plasmon modes of a SP2DEG. 
When the impurities are sufficiently dilute, the collective modes of the disordered system are still given by Eq.~(\ref{eq:condition}), provided that one replaces $\chi^{(0)}_{\sigma}(q,\omega)$ with the disorder-averaged response function. A  suitable expression for the latter was given long ago by Mermin~\cite{mermin_prb_1970}, based on a relaxation-time approximation.  This is actually equivalent, in the diffusive regime, to the sum of impurity ladder diagrams (diffusons)~\cite{Akkermans_and_Montambaux}, but also has the correct behaviour in the high-frequency (collisionless) regime.  The explicit form of the Mermin response function is~\cite{Giuliani_and_Vignale}
\be \label{eq: mermin}
\Pi^{(0)}_{\sigma} (q, \omega) = \frac{ (\omega + i \tau^{-1}) ~\chi^{(0)}_\sigma( q, \omega+  i\tau^{-1})}{\omega + i \tau^{-1} \chi_\sigma^{(0)}(q, \omega+ i \tau^{-1})/\chi_\sigma^{(0)} (q,0)}~,
\ee
where, as before, $\chi^{(0)}_\sigma(q,\omega)$ is the Lindhard response function of the clean SP2DEG~\cite{Giuliani_and_Vignale} and $\tau$ is the elastic transport lifetime that appears in the Drude conductivity of a disordered electronic system. For point scatterers, $\tau$ is given by  $\tau =  \pi/(u_0^2 n_{\rm imp} N_0)$ where $u_0$ is the average disorder strength and $n_{\rm imp}$ is the disorder concentration.

We can solve Eq.~(\ref{eq:condition}) numerically with $\chi^{(0)}_{\sigma} \to \Pi^{(0)}_{\sigma}$ and find how the two modes given by Eqs.~(\ref{eq:chargeplasmon})-(\ref{eq:plasmon_ac}) are affected by static disorder. As a consistency check of our numerical results, we reproduced the results of the clean case, by setting $\tau^{-1}$ = 0 in the above mentioned method---see the solid red and blue curves in Fig.~\ref{fig:fig2}. For the disordered case, we find that each of these modes appears only for wave vectors {\it larger} than a critical value, $q^{*}_{{\rm pl}/{\rm spl}}$---a fact that was first noticed by Giuliani and Quinn~\cite{giuliani_prb_1984} for the charge plasmon. 
As usual, the dispersion of the charge mode and the corresponding critical wave vector $q^{*}_{\rm pl}$ for a finite value of $P$ do not differ significantly from the known analytical results~\cite{giuliani_prb_1984} for $P=0$. 

The spin plasmon mode, whose origin lies in the second zero of the real part of the dielectric function, continues to exist even in the presence of dilute disorder---see Figs.~\ref{fig:fig1}c)-d). 
However, unlike the charge-plasmon mode, the spin-plasmon mode ceases to exist beyond a critical wave vector $q^{\rm max}_{\rm spl}$. 
We now present a detailed analysis of the dependence of the QF of the spin-plasmon mode on various parameters, {\it i.e.}~$\tau, q,P$ and  $r_s$: our main numerical results are reported in Fig.~\ref{fig:fig3}.

In Figs.~\ref{fig:fig3}a)-b), we study the QF of the spin plasmon mode in the $\tau^{-1}-q$ and $\tau^{-1} -P$ plane 
and find that for a given $q$ and $P$, the spin-plasmon mode is sensitive to disorder and vanishes beyond a maximum value of $\tau^{-1}$.  For a given $\tau^{-1}$, the spin-plasmon mode exists in a certain range of values of $q$ and $P$.  The QF of the spin-plasmon mode in the $r_s - P$ plane is shown in Fig.~\ref{fig:fig3}c). We find that for a given parameter set $\tau, q$ and $P$, there is a minimum $r_s$ (maximum density) below (over) which the spin-plasmon mode ceases to exist. Moreover, for a given $r_s$, the spin-plasmon mode exists only in a certain range of polarization values.
Fig.~\ref{fig:fig3}d) displays the QF of the spin-plasmon mode in the $q-P$ plane. Note that $q^{\rm max}_{\rm spl}$ decreases as $P$ increases: this is consistent with the analytical result for the clean case presented 
in Fig.~\ref{fig:fig2}c). Finally, Figs.~\ref{fig:fig3}e)-f) illustrate the QF in the $\tau^{-1}-r_s$ plane and in the $q-r_s$ plane, respectively. In both panels there is a minimum $r_s$, below which the spin-plasmon mode ceases to exist. This happens only in the disordered case and not in the clean limit---see Fig.~\ref{fig:fig2}d). Another feature of the disordered case, namely the existence of a minimum critical wave vector $q^*_{\rm spl}$ below which the spin-plasmon mode ceases to exist, is clearly visible in Fig.~\ref{fig:fig3}f).

\section{``Fermi liquid'' corrections}
\label{sect:Fermiliquidcorrections}

For the sake of clarity and simplicity, the analysis we have carried out in this work has focused on the weak-coupling regime ({\it i.e.}~high-density limit) where the role of exchange and correlations beyond RPA is negligible~\cite{Giuliani_and_Vignale}. Our findings may be substantially altered in the low-density regime. In this case one can refine our calculations by treating exchange and correlations within a generalized random phase approximation~\cite{Giuliani_and_Vignale} and by taking into account the so-called spin-resolved local field factors~\cite{Giuliani_and_Vignale} $G_{\sigma\sigma'}(q,\omega)$. Although some progress has been done over the years~\cite{Giuliani_and_Vignale}, these quantities are still largely unknown. Accurate estimates of the local field factors can be made when the frequency dependence of $G_{\sigma\sigma'}(q,\omega)$, which represents the inertia of the Pauli-Coulomb hole around a reference electron, is neglected. In this case one can indeed use the fluctuation-dissipation theorem~\cite{Giuliani_and_Vignale} in conjunction with accurate knowledge~\cite{gorigiorgi_prb_2004} of the spin-resolved pair distribution function $g_{\sigma\sigma'}(r)$. This approach has been taken, for example, in Refs.~\onlinecite{asgari_prb_2006,abedinpour_arxiv_2012} and can be generalized to the case of a SP2DEG. 
 
It is known~\cite{Giuliani_and_Vignale} that, in the static limit, the local field factors $G_{\sigma\sigma'}(q,0)$ vanish linearly in $q$. At the same time, the bare Coulomb interaction $v_q$ diverges as $q^{-1}$.  In the small-$q$ regime, we can therefore characterize the strength of the local field correction via the ``Landau-like parameters''
\be
F_{\sigma\sigma'}\equiv -N_0 \lim_{q\to 0} v_qG_{\sigma\sigma'}(q,0)~.
\ee
These parameters control the many-body corrections to thermodynamic response functions such as the compressibility, the spin susceptibility, and the mixed spin-density susceptibility.
The dispersion and damping of the plasmons are now determined by the zeroes of the generalized dielectric function
\ber
 \epsilon(q,\omega)&=& [1-v_{\uparrow\uparrow}(q)\chi^{(0)}_{\uparrow}(q,\omega)][1-v_{\downarrow\downarrow}(q)\chi^{(0)}_{\downarrow}(q,\omega)] \nn\\
 &-&[v_{\uparrow\downarrow}(q)]^2\chi^{(0)}_{\uparrow}(q,\omega)\chi^{(0)}_{\downarrow}(q,\omega)~,
 \eer
where $v_{\sigma\sigma'}(q)\equiv v_q[1-G_{\sigma\sigma'}(q,0)]$.
 This reduces to the RPA dielectric function introduced in Sect.~\ref{sect:RPAtheory} when the local field factors are set to zero.
 It is now a matter of straightforward algebra to calculate the local field corrections to the quantities calculated in Sect.~\ref{sect:RPAtheory}. 
Working in the long wave-length limit and incorporating these corrections we find 
\be \label{eq:omegaFL}
\omega_{\rm spl}(q\to 0)= c_{\rm s} q  \frac{1+ F_{\rm a}/2}{\sqrt{1 + 2 F_{\rm a}/3 }}~,
\ee 
 and
\be \label{eq:QFL}
Q_{\rm spl}=\frac{(3+2F_{\rm a}) \sqrt{7p-1 + F_{\rm a} (6p-1) - F_{\rm a}^2 (1-p)}}{\sqrt{1-P} (1+F_{\rm a})^2}~,
\ee
where $c_{\rm s}$ has been defined in Eq.~(\ref{eq:velocity}) and $F_{\rm a} \equiv F_{\uparrow\uparrow}+F_{\downarrow\downarrow}-2F_{\uparrow\downarrow}$. 
In the limit $F_{\sigma\sigma'} \ll 1$, and working to first order in these corrections,  Eqs.~\eqref{eq:omegaFL}-\eqref{eq:QFL}, reduce to
\be
\omega_{\rm spl}(q\to 0)= c_{\rm s} q \left(1+\frac{F_{\rm a}}{6}\right)~,
\ee
and 
\be
Q_{\rm spl}=  3 \sqrt{\frac{7 p-1}{1 + p}} \left(1 - F_{\rm a} \frac{19 p - 1}{7 p - 1} \right)~.
\ee 

The frequency dependence of the local field factors, however, can be very important in the spin channel, in the low-density (large $r_s$) regime~\cite{Giuliani_and_Vignale}. 
A careful account of many-body corrections to $\omega_{\rm spl}(q\to 0)$ and 
$Q_{\rm spl}(q \to 0)$ beyond the RPA is well beyond the scope of the present Article and is left for future work.

\section{Summary and conclusions}
\label{sect:summary}

In summary, we have discovered that a partially spin-polarized two-dimensional electron gas supports a spin-plasmon mode with a gapless acoustic dispersion. We have fully characterized the mode energy and its linewidth with and without static disorder within the random phase approximation. The spin-plasmon mode discussed in this Article can be probed directly with inelastic light scattering~\cite{pinczuk, ILS} in a two-dimensional electron gas hosted {\it e.g.} in a high-quality modulation-doped GaAs quantum well subject to a Zeeman field perpendicular to the growth direction.  

We note here that there is no mechanism of spin relaxation (spin flip) in our model. The inclusion of a spin relaxation mechanism ({\it e.g.}~via spin-orbit coupling, or exchange correlations, or spin-flip scattering) in a ferromagnet is known to lead to a gapped finite frequency spin mode (Stoner excitations) in the homogeneous limit ($q \to 0$) \cite{Yoshida}. However, this finite frequency hydrodynamic mode is completely different in nature from our collisionless spin plasmon, in which a time-dependent longitudinal spin polarization is created not by spin flip processes but by different motions of up and down spin electrons. Here we expect the addition of spin-orbit interaction to modify the damping of the mode and reduce the quality factors.

We now describe the implications of this stable spin plasmon in the SP2DEG for spin-wave coupling between oscillating nanomagnets driven by spin torque (spin-torque oscillators). These oscillators are characterized by a current-dependent resonant frequency $\omega_r$, and interaction and phase locking has been demonstrated through the emission and absorption of spin waves in the intervening material\cite{kaka2005,tehrani2005}. The SP2DEG provides an intervening material with a response that is  tunable through control of the spin polarization. For a sufficiently large value of $P$, and for a Fermi energy such that $\omega_{\rm spl}(q) = \omega_r$, for a region of the spin plasmon curve that is outside the minority-spin single-particle continuum, 
these spin plasmons will be excited and propagate  over distances  in excess of the minimum decay length $l$. The minimum decay length, $l \approx 2 \pi Q_{\rm spl}/q$ and is approximately  $2 \mu$m for typical 2DEG electron densities of $10^{15} $m$^{-2}$ (in GaAs this corresponds to $r_s = 1.8$),  and for other parameters given by the maxima of $\gamma_{\rm spl}$ curve (filled red circles) in Fig.~\ref{fig:fig2}b), which corresponds to the minimum decay time for the spin-plasmon mode. 
The coupling between the spin torque oscillators can be controlled by changing the value of $P$ or the local density of electrons (by applying local gate voltages) such that  $\omega_{\rm spl} (q) \neq \omega_r$. 
If the value of $P$ is reduced such that $\omega_{\rm spl} (q) = \omega_r$ lies within the 
minority-spin single-particle continuum, then the propagation distance of these spin plasmons will be only a couple of Fermi wave-lengths.
Thus it is possible to tune the efficiency of coupling between spin torque oscillators, and affect their phase locking, simply by changing the polarization of an intervening SP2DEG that couples them. 

Note that the polarization of a spin-torque oscillator is transverse, and precesses around a bias magnetic field. In contrast the spin plasmons oscillate parallel to the spin polarization in the SP2DEG. However, if the spin polarization in the SP2DEG is established through spin injection rather than an applied magnetic field, it can be established perpendicular to the bias magnetic field applied to the spin-torque oscillator, permitting coupling between the magnetization oscillations of the spin-torque oscillator and the stable spin plasmon modes of the SP2DEG.

\acknowledgments 
This work was supported in part (M.E.F.) by C-SPIN, one of six centers of STARnet, a Semiconductor Research Corporation program, sponsored by MARCO and DARPA. We also acknowledge financial support by the EU FP7 Programme under Grant Agreement No. 215368-SEMISPINNET (A.A. and M.P.), the Italian Ministry of Education, University, and Research (MIUR) through the program ÒFIRB - Futuro in Ricerca 2010Ó Grant No. RBFR10M5BT (M.P.), and an ARO MURI (M.E.F. and G.V.). A.A. gratefully acknowledges funding from the INSPIRE faculty fellowship by DST  (Govt. of India), and from the Faculty Initiation Grant by IIT Kanpur, India.

\end{document}